\documentclass[letterpaper, 10 pt, journal, twoside]{IEEEtran}
\usepackage{amsmath,amsfonts}
\usepackage{algorithmic}
\usepackage{algorithm}
\usepackage{array}
\usepackage[caption=false,font=normalsize,labelfont=sf,textfont=sf]{subfig}
\usepackage{textcomp}
\usepackage{stfloats}
\usepackage{url}
\usepackage{verbatim}
\usepackage{graphicx}
\usepackage{cite}
\usepackage{xcolor}
\usepackage{bm}
\usepackage[]{units}
\usepackage{mathrsfs}
\usepackage{mathtools}
\usepackage{subfig}
\usepackage{wasysym} % for diameter symbol
\usepackage{hyperref}
\usepackage{xfrac} % required for diagonal fraction 

\DeclareMathSizes{10}{10}{6}{1}

\def\<{ \begin{bmatrix} }
\def\>{ \end{bmatrix} }

\renewcommand{\d}{\mathrm{d}} 			% differential d, used e.g. in d/dt
\renewcommand{\b}{\bm{b}}     			% magnetic field vector
\renewcommand{\ss}{\mathrm{ss}} 		% steady-state index
\renewcommand{\phi}{\varphi} 		    % to avoid any accidental phi's
\newcommand{\current}{\bm{i}} 		    % electrical current i 
\renewcommand{\L}{\bm{\mathrm{L}}} 		% ILC update L -> u
\renewcommand{\deg}{\text{\textdegree}}   % [deg] (circle)
\newcommand{\reg}{\textregistered\,}	% registered symbol
\newcommand{\SP}{\mathrm{SP}} 	 		% setpoint index
\newcommand{\emm}{\mathrm{m}}           % index for magnet
\newcommand{\A}{\bm{\mathrm{A}}}		% system matrix A
\newcommand{\B}{\bm{\mathrm{B}}}		% input matrix B
\newcommand{\C}{\bm{\mathrm{C}}}		% measurement matrix C
\newcommand{\K}{\bm{\mathrm{K}}}		% state feedback K
\newcommand{\x}{\bm{x}} 		 		% state vector
\newcommand{\Zero}{\bm{0}} 		 		% zeros matrix 
\newcommand{\I}{\bm{\mathrm{I}}} 		% identity matrix I
\newcommand{\Act}{\bm{\mathcal{A}}} 	% actuation matrix 
\newcommand{\PsiB}{\mathrm{\Psi}_b}     % allocation u -> b
\newcommand{\Pilc}{\bm{\mathrm{P}}} 	% lifted system matrix for ILC
\newcommand{\Dilc}{\bm{\mathrm{N}}} 	% derivative operator
\newcommand{\Q}{\bm{\mathrm{Q}}} 		% ILC update Q -> e
\newcommand{\T}{\top} 			% transpose
\newcommand{\mb}{|\tilde{\bm{m}}||\bm{b}|} % product |m||b|

\newlength{\varminus}
\begin{document}

\title{Dynamic Electromagnetic Navigation}

\author{Jasan Zughaibi, Bradley J. Nelson, and Michael Muehlebach\vspace{-2.9mm}
\thanks{Manuscript received: January 21, 2025; Revised March 25, 2025; Accepted April 09, 2025.}%Use only for final RAL version
\thanks{This paper was recommended for publication by Editor Jessica Burgner-Kahrs upon evaluation of the Associate Editor and Reviewers' comments. This work was supported by the Swiss National Science Foundation under Grant IZLCZ0\_206033, the Max Planck ETH Center for Learning Systems, the German Research Foundation, and the Branco Weiss Fellowship. }
\thanks{Jasan Zughaibi and Bradley J. Nelson are with the Multi-Scale Robotics Lab, ETH Zurich, Switzerland  {\tt\footnotesize zjasan@ethz.ch, bnelson@ethz.ch}}
\thanks{Michael Muehlebach is with the Learning and Dynamical Systems Group, Max Planck Institute for Intelligent Systems Tübingen, Germany  {\tt\footnotesize michael.muehlebach@tuebingen.mpg.de}}
\thanks{Digital Object Identifier (DOI): see top of this page.}
}

% The paper headers
%\markboth{Journal of \LaTeX\ Class Files,~Vol.~14, No.~8, August~2021}%
%{Shell \MakeLowercase{\textit{et al.}}: A Sample Article Using IEEEtran.cls for IEEE Journals}

%\IEEEpubid{0000--0000/00\$00.00~\copyright~2021 IEEE}
% Remember, if you use this you must call \IEEEpubidadjcol in the second
% column for its text to clear the IEEEpubid mark.

\maketitle

\begin{abstract}
	Magnetic navigation offers wireless control over magnetic objects, which has important medical applications, such as targeted drug delivery and minimally invasive surgery. Magnetic navigation systems are categorized into systems using permanent magnets and systems based on electromagnets. Electromagnetic Navigation Systems (eMNSs) are believed to have a superior actuation bandwidth, facilitating trajectory tracking and disturbance rejection. This greatly expands the range of potential medical applications and includes even dynamic environments as encountered in cardiovascular interventions. To showcase the dynamic capabilities of eMNSs, we successfully stabilize a (non-magnetic) inverted pendulum on the tip of a magnetically driven arm. Our approach employs a model-based framework that leverages Lagrangian mechanics to capture the interaction between the mechanical dynamics and the magnetic field. Using system identification, we estimate unknown parameters, the actuation bandwidth, and characterize the system's nonlinearity. To explore the limits of electromagnetic navigation and evaluate its scalability, we characterize the electrical system dynamics and perform reference measurements on a clinical-scale eMNS, affirming that the proposed dynamic control methodologies effectively translate to larger coil configurations. A state-feedback controller stabilizes the inherently unstable pendulum, and an iterative learning control scheme enables accurate tracking of non-equilibrium trajectories. Furthermore, to understand structural limitations of our control strategy, we analyze the influence of magnetic field gradients on the motion of the system. To our knowledge, this is the first demonstration to stabilize a 3D inverted pendulum through electromagnetic navigation.
\end{abstract}

\section{Introduction}
\IEEEPARstart{M}{agnetic} navigation - often referred to as magnetic actuation - describes the use of magnetic fields to exert wireless control over the spatial orientation and positioning of magnetic objects \cite{petruska2020magMethodsRobot}. These objects can vary considerably in size, spanning scales from nanometers to centimeters. There is considerable promise in utilizing this technology for advanced medical applications \cite{zhao2022tele_mag}, \cite{hongsoo2020vascular}. This includes the deployment of biocompatible micro- and nanorobots for targeted drug delivery and magnetic continuum robots, such as catheters and guidewires, for minimally invasive surgeries \cite{gao2017micronano}, \cite{zhao2019ferromagSCR}.
% e.g. GI tract, vasculature, brain, eye, 
% cardiac ablations and neurovascular interventions

\begin{figure}
	\centering
	\includegraphics[trim=52mm 22mm 13mm 89mm, clip, width=0.71\columnwidth]{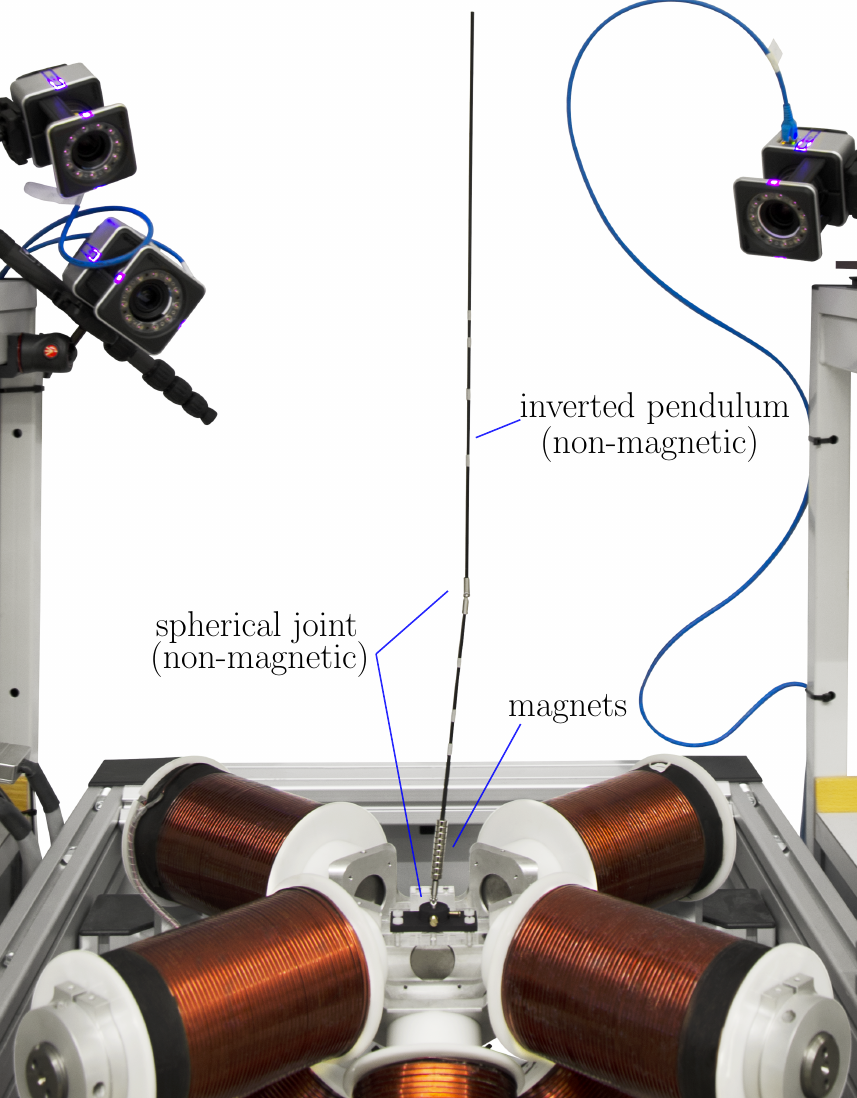}
	\vspace{-0.3cm} % [left bottom right top]
	\caption{3D inverted pendulum balanced using the OctoMag electromagnetic navigation system. The system includes an arm driven by the external magnetic field, and a non-magnetic pendulum, free to rotate spherically. Both arms are connected through a (non-magnetic) spherical joint. }
	\label{fig:octomag_3D_pend}
    \vspace{-7mm}
\end{figure}
A system that is used to control a magnetic object in space is often referred to as a magnetic navigation system. Magnetic navigation systems can be broadly categorized into systems that rely on the motion of permanent magnets and systems that generate magnetic fields using a set of electromagnets, so-called electro-Magnetic Navigation Systems (eMNSs). Systems such as the Stereotaxis Genesis\reg and Niobe\reg use robotic arms manipulating permanent magnets with a mass of the order of several hundred kilograms \cite{kladko2023magnetosurgery}. The high inertia restricts the physical motion and
results in low actuation bandwidth \cite{zhang2020magnetic}. However, there are also established systems employing robotic arms with much lighter permanent magnets, with some designs featuring magnets weighing \unit[1.5]{kg} or \unit[7]{kg} \cite{valdastri2010rms_1p5kg}, \cite{zhao2022tele_mag}.
The potential impact of these lighter systems on actuation bandwidth remains an unexplored area of research, as no study has investigated or directly compared the actuation bandwidth of eMNSs and systems that utilize permanent magnets. Nonetheless, eMNSs are generally understood to provide a higher bandwidth, which offers several benefits, including improved trajectory tracking and superior disturbance rejection. This becomes crucial in dynamic environments encountered in cardiovascular interventions, such as cardiac ablation therapy, where precise movement is required \cite{howe2011cardiac_ctrl}.
% enabling more precise and quicker adjustments during procedures.

%Achieving accurate control with robotically
%actuated permanent magnets [4] is challenging, due largely to
%the high inertia related to the movements of the large EPM and
%serial manipulator, compared to current flow

In this letter, we explore the dynamic capabilities of eMNSs by stabilizing a (non-magnetic) inverted pendulum on the tip of a magnetically driven arm (see Fig. \ref{fig:octomag_3D_pend}). A video of the system in operation is available here: \href{https://youtu.be/FU6rylNam28}{\texttt{https://youtu.be/FU6rylNam28}}. In the field of control theory, the inverted pendulum is a classical and extensively studied system with numerous practical implementations \cite{gluck2013triplePend}, \cite{astromFuruta2000furutaPend}, \cite{muehlebach2017cubli}. It serves as a benchmark for novel control architectures and is of high educational value \cite{boubaker2012invPendEduBenchmark}. Due to its inherent instability, a successful stabilization necessitates a sufficiently high actuation bandwidth. 

Beyond merely demonstrating the dynamic potentials of eMNSs, the solution to this challenging control problem offers valuable insights into magnetic control strategies, particularly in the presence of the inherent strong nonlinearities of magnetic fields. Such insights could be adapted for a variety of tasks, making it an intriguing testbed and benchmark problem for the study of novel magnetic control algorithms. To the best of our knowledge, this study represents the first effort to stabilize a 3D inverted pendulum using magnetic navigation. While the inverted pendulum scenario is not clinically relevant in itself, our methods are broadly applicable and can be readily adapted to control continuum robots or catheters. Crucially, this work encourages the development of dynamic control algorithms given that the majority of literature on electromagnetic navigation relies on feedforward control and quasi-static modeling. By demonstrating robust disturbance-rejection capabilities of our algorithms, we also aim to motivate the development of real-time state-estimation techniques suitable for in-vivo applications, as reliable state information is fundamental to enabling dynamic feedback control.

To investigate the fundamental bandwidth limitations of eMNSs, we characterize the dynamics of the electrical system by performing system identification experiments. We compare the frequency responses of the setup used in this work to a clinical-scale eMNS, offering insights into the scalability of our dynamic control approach and examining its potential impact on medical applications.

Our objective is to control the magnetically driven arm using an external magnetic field, such that the (non-magnetic) pendulum on top of the arm maintains in its upright equilibrium position. To produce the magnetic field, we utilize the OctoMag system \cite{kummer2010octomag}, which comprises eight electromagnets/coils. It should be noted that, theoretically, achieving this task could be possible with fewer than eight coils. This is because we primarily exploit the orientation of the magnetic field vector to exert torques, assuming a homogeneous magnetic field in the proximity of the center of the workspace \cite{petruska2015MinimumBounds}. 

\subsection{Methodology}
We use a model-based framework for designing the control algorithms. First, the dynamics describing the interaction between the magnetic field and the pendulum system are captured through Lagrangian mechanics. We then perform system identification experiments in order to estimate unknown parameters and characterize the electrical dynamics of the system. Our identification procedure enables not only a precise estimation of the mechanical/electrical system's frequency response and their actuation bandwidths, but also provides an uncertainty estimate that captures the system's nonlinearity.  

We then design a state-feedback controller based on our identified dynamical model that is capable of stabilizing pendulums ranging from 20 to \unit[40]{cm} in length. In addition, we introduce model-based techniques that offer online compensation for errors arising from the calibration of the magnetic field and the calibration of the angle measurement system. 

Our work goes beyond the mere stabilization at equilibrium and also includes reference tracking controllers. These are capable of following dynamic non-equilibrium motions with the magnetic arm, while balancing the pole in its upright position. To that extent, we augment the control system with an Iterative Learning Control (ILC) scheme that leverages the identified dynamic model to compute a feedforward correction signal based on past tracking information. By iterating the same task, the ILC learns to systematically compensate for repetitive tracking errors. Finally, we discuss structural limitations of our control strategy by examining the influence of magnetic field gradients on the motion of the system.  

\subsection{Related Work}
A common assumption in the field of magnetic navigation is that dynamic effects can be neglected for modeling and control, resulting in quasi-static models \cite{edelmann2017mag_ctrl}, \cite{cenk2014_3Dmodeling}. Although this assumption may be adequate for a variety of medical applications, there is great potential for utilizing dynamic models and dynamic control algorithms. 

The authors of \cite{valdastri2019rmn_levitation_dyn} derive a dynamic model that captures the interaction of an external, robotically steered, permanent magnet and a magnetic endoscope. This model serves as the foundation for synthesizing dynamic control strategies, enabling the magnetic tip of the tethered endoscope to be elevated while navigating along curved paths. In a complementary study, a similar experiment is demonstrated utilizing a model predictive controller. This predictive controller accounts for the dynamics and system constraints by formulating an optimal control problem that is solved in explicit form \cite{valdastri2019MPC}.  

In \cite{cavusoglu2018sysidMRIcatheter}, the dynamics of an MRI-guided catheter are identified and analyzed. The authors approximate the dynamics of the catheter using a 2D linear dynamical black-box model, identified by exciting the system with a chirp signal. The work of \cite{hamal2014robustMicro} derives a dynamic model of a microrobot controlled by an eMNS. Parameter uncertainties are incorporated into the model, leading to the synthesis of an $\mathcal{H}_\infty$ robust control algorithm. The authors of \cite{sitti2009dynamicMicro} develop a dynamic model of a microrobot, actuated by a pair of Helmholtz coils. The model is utilized as a simulation tool to understand the influence of design parameters on the microrobot's motion behavior.
% further?
% DillerSetti dynamics: "Independent control of multiple magnetic microrobots in three dimensions"
% petruska: "Observed Control of Magnetic Continuum Devices"

The application of ILC to improve the tracking performance of inverted pendulum systems has been explored by several researchers in the field. The work of \cite{hehn2013ILCpend} employs a frequency-domain ILC scheme to enhance the tracking accuracy of a quadrocopter while maintaining the stability of an inverted pendulum. Their approach leverages the periodicity of the trajectories to decompose the correction and error signal using a Fourier series, resulting in an ILC algorithm that has low computational cost. A more general approach that parametrizes inputs and state trajectories with basis functions has been proposed in \cite{carlo2020mpc}, where similar robotic testbeds have been used. In \cite{petriu2009ILC_pend}, the potential of proportional-derivative-type (PD) ILC schemes is explored, implemented in both serial and parallel configurations, aiming to enhance the tracking performance of an inverted pendulum system mounted on a cart, while metaheuristic-based tuning of PD-type ILC for other applications is discussed in \cite{precup2024metaheuristic}.

\subsection{Outline}
The remainder of this document is organized as follows: Sec. \ref{sec:Experimental_Platform} details the experimental setup and defines all relevant quantities, including control inputs, parameters, and coordinates. The modeling process, parameter estimation, and analysis of the electrical system dynamics are presented in Sec. \ref{sec:Modeling}. Feedback controller synthesis and online compensation algorithms that remove calibration errors from the actuation and sensors are presented in Sec. \ref{sec:State_Feedback_Control}. The synthesis of the ILC, which greatly improves the trajectory tracking capabilities, is discussed in Sec. \ref{sec:Iterative_Learning_Control}. Finally, a conclusion is drawn in Sec. \ref{sec:Conclusion}.

\section{Experimental Platform}
\label{sec:Experimental_Platform}
The following section describes the experimental setup used for stabilizing the inverted pendulum. Furthermore, we define the control inputs and their relation to the electrical currents of the OctoMag system.

\begin{figure}
	\centering
	\includegraphics[trim={6cm 17.8cm 3cm 3.4cm},clip, scale=0.81]{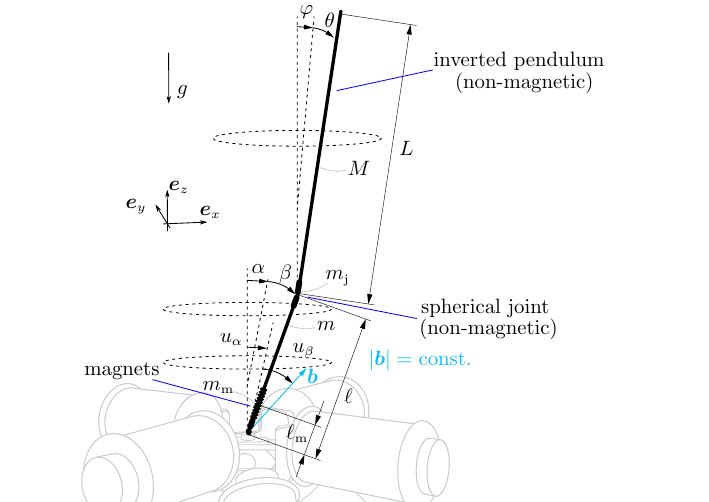} % left, bottom, right, top
	\vspace{-20pt}
	\caption{The figure illustrates the parametrization of orientations, describing rotations with respect to the inertial frame of the motion capture system. Both the actuator and the inverted pendulum possess two kinematic degrees of freedom. These are parametrized by the generalized coordinates $\alpha$ and $\beta$ for the actuator, and $\varphi$ and $\theta$ for the inverted pendulum. The control input to the system is the orientation of the magnetic field vector $\bm{b}$, parametrized by the angles $u_\alpha, u_\beta$. Rotations described by $u_\alpha, \alpha,$ and $\varphi$ are with respect to the inertial $\bm{e}_y$ axis, whereas $u_\beta, \beta,$ and $\theta$ correspond to the inertial $\bm{e}_x$ axis. We demonstrate pendulum stabilization of lengths between $L=\unit[20-40]{cm}$, with an actuator length of $\ell=\unit[22]{cm}$. The distance between the pivot point and the magnet center is denoted by $\ell_\emm$. The pendulum, actuator, joint, and magnet masses are denoted by $M$, $m$, $m_\mathrm{j}$, and $m_\emm$, respectively. }
	\label{fig:notation}
    \vspace{-5mm}
\end{figure}

The pendulum assembly consists of two commercially available rods made from carbon-fiber reinforced thermoplastics. We refer to the lower arm as the \textit{actuator} and the upper arm as the (non-magnetic) \textit{pendulum}. Ten axially magnetized permanent magnets (Magnetkontor\reg R-08-03-04-N3-N, NdFeB N45,  \diameter8 (3) x 4 mm) are mounted proximally to the actuator's pivot point, as illustrated in Fig. \ref{fig:notation}. The actuator is connected to a base plate using a non-magnetic u-joint (303 stainless steel, McMaster-Carr\reg 60075K75), that allows the actuator to rotate spherically. We use the same type of u-joint to connect the actuator with the pendulum, resulting in a total of four (angular) degrees of freedom. The actuator's orientation is defined by the angles, $\alpha, \beta$, while the orientation of the pendulum is described by $\varphi, \theta$, as illustrated in Fig. \ref{fig:notation}. 

Reflective marker stripes are attached at the actuator and the pendulum rod which allows us to retrieve all angles at a frequency of \unit[100]{Hz} using a motion capture system.

%All geometric and mass-related parameters are listed in Tab. \ref{tab:geom_mass_related_prm}, their definitions are provided in Fig. \ref{fig:notation}.  
%
%
%\begin{table}[h]
%	\centering
%	\caption{Summary of Parameters. \mynotes{check consistency of mass, density, length pendulums}}
%	\begin{tabular}{c|c|c|c|c|c|c}
%		 \( L \) & \( \ell \) & \( l_\emm \) & \( m \) & \( M \) & \( m_\mathrm{j} \) & \( m_\emm \) \\
%		\hline
%		 \unit[405]{mm} & \unit[218]{mm} & \unit[38]{mm} & \unit[2.4]{g} & \unit[4.4]{g} & \unit[2.0]{g} & \unit[12.7]{g} \\
%	\end{tabular}
%	\label{tab:geom_mass_related_prm}
%\end{table}

% Table with parameters ? 
% L = 405mm 
% l = 200mm 
% l_m
% m = 0.0024 kg 
% M = 0.0044 kg 
% m_joint = gen.g2kg(2.0); 
% m_magnet = gen.g2kg(1.27);
% obj.l_mag_nom   = gen.mm2m(38.0); 
% obj.l_stick_nom = gen.mm2m(218);    % [m]      % length of the stick

% (PA 12 jet fusion, as all 3D printed parts)
% mention that shaft of u-joint is slightly magnetic?

\subsection{Magnetic Field Allocation}
We define the control inputs to the system as the angles, $u_\alpha, u_\beta$, that parametrize the orientation of the magnetic field vector, $\bm{b} \in \mathbb{R}^3$, in an inertial coordinate frame (see Fig. \ref{fig:notation}). The relation between $u_\alpha, u_\beta$ and $\bm{b}$ is given by 
\begin{align*}
	\PsiB: (u_\alpha, u_\beta, |\b|) \mapsto \b:\, \< b_x \\ b_y \\ b_z \> = |\bm{b}| \< \sin(u_\alpha)\cos(u_\beta) \\  -\sin(u_\beta) \\ \cos(u_\alpha)\cos(u_\beta) \>, %\label{eq:mag_field_allocation}
\end{align*}
where $|\bm{b}|$ is a free parameter that can be specified by the user. We keep the magnitude of the magnetic field vector constant at all times, specifically $|\bm{b}| = \unit[35]{mT}$. From a control systems perspective, it is advantageous to use the angles, $u_\alpha, u_\beta$, as control inputs, since they are directly connected to the angular deflections, $\alpha, \beta$, of the actuator. This allows us to consider the dynamics as decoupled, simplifying our control strategy.  

For eMNSs it is commonly assumed to rely on a linear relation between electrical currents, $\current \in \mathbb{R}^8$, the field, $\bm{b}$, and its gradient, $\bm{g}=\left(\begin{array}{lllll}
	\frac{\partial b_x}{\partial x} & \frac{\partial b_x}{\partial y} & \frac{\partial b_x}{\partial z} & \frac{\partial b_y}{\partial y} & \frac{\partial b_y}{\partial z}
\end{array}\right)^\T$, i.e.
\begin{align}
 \< \bm{b} \\ \bm{g} \> = \bm{\mathcal{A}}\,\current \qquad \Rightarrow \qquad \current_\SP = \Act^\dagger \< \b_\SP\\ \Zero \>, \label{eq:currents_2_magfield}
\end{align}
where $\bm{\mathcal{A}}\in \mathbb{R}^{8 \times 8}$ is the so-called actuation matrix at the center of the magnetic workspace \cite{quentin2023workspace}. Notice, that five gradient terms uniquely describe the gradient, as there are four constraints resulting from Maxwell's equations (static and absence of free currents), namely $\nabla \times \b = 0$ and $\nabla \cdot \b = 0$. In general, the matrix $\bm{\mathcal{A}}$ depends on the configuration of the coils and can be determined through a calibration procedure as described in \cite{petruska2020magMethodsRobot}. Given a desired field, $\bm{b}_\SP$, the desired currents can be determined using the Moore-Penrose pseudoinverse, denoted by $\dagger$ (see Eq. \eqref{eq:currents_2_magfield}). By setting $\bm{g}_\SP = \bm{0}$ we achieve (approximately) a homogeneous, i.e. gradient-free, magnetic field in the vicinity of the center of the magnetic workspace. 
\section{Modeling}
\label{sec:Modeling}
In this section, we describe the modeling and parameter estimation processes for the combined actuator-pendulum system. The procedure unfolds in two stages. First, we develop an analytical model capturing the system’s dynamics and its interaction with the magnetic field. Next, we conduct system identification experiments to obtain data-driven estimates of unknown model parameters. Finally, we characterize the electrical dynamics to investigate the scalability of our dynamic control approach to clinical-scale eMNSs.

% assuming decoupled dynamics, model requierments sufficiently descriptive for controller synthesis, assuming that uncertainties can be compensated using feedback control and a learning control schemes (see sec)

\subsection{Analytical Model}
Linearizing the nonlinear dynamics of a 3D inverted pendulum around its upright equilibrium yields two decoupled 2D linear models. This decoupling suggests that the full 3D behavior can be effectively understood by examining these individual 2D systems separately, given that the angles are sufficiently small. Consequently, for clarity and brevity, we focus our attention on deriving the dynamics for one of these 2D planes, specifically the $(\alpha, \varphi)$-plane, using Lagrangian mechanics with $(\alpha, \varphi)$ as generalized coordinates. We account for the system's interaction with the external magnetic field by incorporating its potential energy in the Lagrangian. The potential energy of a magnetic dipole in an external (2D) field is given by $U_{\mathrm{m}}= - |\bm{\tilde{m}}||\bm{b}| \cos(u_\alpha - \alpha),$ where $\bm{\tilde{m}}$ (in $\mathrm{Am^2}$) denotes the magnetic dipole moment. The Lagrangian of the system reads as, $\mathscr{L} = T - U$, where $T$ and $U$ are the kinetic and potential energy, respectively:
\begin{align*}
	T &=\frac{1}{2}\left(J+M \ell^2\right) \dot{\alpha}^2+\frac{1}{8} M L^2 \dot{\varphi}^2+\frac{1}{2} M \ell L \dot{\alpha} \dot{\varphi} \cos (\alpha-\varphi) \\
	U &= (\eta+M \ell) g \cos \alpha+M g \frac{L}{2} \cos \varphi+U_{\mathrm{m }},
\end{align*}
where we introduce the abbreviations, $J$ and $\eta$, defined as
\begin{align*}
	J \coloneqq m_{\mathrm{m}} \ell_m^2 + m \frac{\ell^2}{4} + m_\mathrm{j} \ell^2, \qquad \eta \coloneqq m_\mathrm{m} \ell_\mathrm{m} + m \frac{\ell}{2} + m_\mathrm{j} \ell.
\end{align*}   
The nonlinear equations of motion can be derived by applying the Lagrange formalism, namely 
\begin{align*}	
	\frac{\d }{\d t} \biggl( \frac{\partial \mathscr{L}}{\partial \dot{\alpha}}\biggr) - \frac{\partial \mathscr{L}}{\partial \alpha} = Q_\alpha^{\mathrm{nc}},  \qquad
	\frac{\d }{\d t} \biggl( \frac{\partial \mathscr{L}}{\partial \dot{\varphi}}\biggr) - \frac{\partial \mathscr{L}}{\partial \varphi} = Q_\varphi^{\mathrm{nc}}, 
 \end{align*}
where $Q_\alpha^{\mathrm{nc}}, Q_\varphi^{\mathrm{nc}}$ represent the non-conservative generalized forces, which primarily arise from the friction in the u-joint connecting the actuator to its base and between the actuator and the pendulum. We assume that the damping torque is directly proportional to the angular velocities, i.e., $Q_\alpha^{\mathrm{nc}}=-d\dot{\alpha} -d(\dot{\alpha} - \dot{\varphi})$, $Q_\varphi^{\mathrm{nc}} = -d (\dot{\varphi} - \dot{\alpha})$. Linearizing the nonlinear dynamics around the (upright) equilibrium $(\alpha, \varphi, u_\alpha) = (0,0,0)$, results in
\begin{align}
\bm{\mathrm{M}}\< \ddot{\alpha} \\ \ddot{\varphi} \>+\bm{\mathrm{D}}\< \dot{\alpha} \\ \dot{\varphi} \>+ \bm{\mathrm{K}}\<\alpha \\ \varphi \> = \bm{w} u_\alpha, \label{eq:lin_eom}
\end{align}
\vspace{-10pt}
\begin{align*}
	\begin{aligned}
		\bm{\mathrm{M}} &= \<
		J+M \ell^2 & \frac{1}{2} M \ell L \\
		\frac{1}{2} M \ell L & \frac{1}{4} M L^2
		\> \\
		\bm{\mathrm{K}} &= \<
		-(\eta +M \ell) g + |\bm{\tilde{m}}| |\bm{b}| & 0 \\
		0 & -Mg \frac{L}{2}
		\>
	\end{aligned}
	& \hspace{3mm}
	\begin{aligned}
		\bm{\mathrm{D}} &= \<
		2d & -d \\
		-d & d
		\> \\
		\bm{w} &= \<
		|\bm{\tilde{m}}| |\bm{b}| \\
		0
		\>.
	\end{aligned}
\end{align*}
The term $|\bm{\tilde{m}}||\bm{b}|$ in the stiffness matrix, $\bm{\mathrm{K}}$, can be interpreted as the magnetic field providing a stiffness-like effect on the actuator, causing the actuator to stably align with the external field around some steady-state value, given that $|\bm{b}|$ is sufficiently large. We rewrite \eqref{eq:lin_eom} as a first order state space model, by introducing the state $\x_\alpha \coloneqq \< \alpha & \varphi & \dot{\alpha} & \dot{\varphi} \>^\T$, i.e. 
\begin{align}
	\dot{\x}_\alpha &= \< \bm{0} & \bm{\mathrm{I}} \\ -\bm{\mathrm{M}}^{-1} \bm{\mathrm{K}} & -\bm{\mathrm{M}}^{-1}\bm{\mathrm{D}} \>   \x_\alpha + \< 0 \\ \bm{\mathrm{M}}^{-1} \bm{w}\>u_\alpha.
    \label{eq:state_space_cont_time}
\end{align} 
Similarly, we define the state for the $(\beta, \theta)$-plane as $\x_\beta \coloneqq \< \beta & \theta & \dot{\beta} & \dot{\theta}  \>^\T$. For the sake of brevity, let $\x$ represent either $\x_\alpha$ or $\x_\beta$ for the remainder of this document. Note that \eqref{eq:state_space_cont_time} captures the bidirectional coupling between the actuator and the inverted pendulum dynamics. Although such coupling is often neglected in standard inverted pendulum models (see e.g.\cite{gluck2013triplePend}), we include it here because the mass of the pendulum is on the same order of magnitude as that of the actuator.

\subsection{Parameter Estimation and System Identification}
In the linearized equations of motion \eqref{eq:lin_eom}, the only unknown parameters are the damping coefficient $d$ and the product $\mb$. The mass terms in \eqref{eq:lin_eom} are obtained using a weighing scale, while the geometric parameters are derived from a computer-aided design model. We identified two parameter estimation methods sufficient for stabilizing the pendulum
\begin{enumerate}
    \item[i)] We set $d = 0$ and estimate $\lvert \tilde{\bm{m}} \rvert$ from material properties, i.e. $\lvert \tilde{\bm{m}} \rvert = b_r V/\mu_0$, where $b_r$ is the remanence, $V$ is the magnetic volume, and $\mu_0$ is the vacuum magnetic permeability. The magnitude of the field vector $\lvert \bm{b} \rvert$ is approximately known from the eMNS calibration.
    \item[ii)] Both $d$ and $\mb$ are estimated from system identification experiments. This approach compensates for spatial variations of $|\bm{b}|$ across the magnetic volume (see Sec.~\ref{sec:Iterative_Learning_Control} for a detailed analysis), thereby having a corrective effect on the product $\mb$. Because the inverted pendulum is inherently unstable, these experiments must be performed without the pendulum attached.
\end{enumerate}

The remainder of this paragraph is closely inspired by the system identification methodology presented in \cite{hao2022learning}. We perform a system identification in the frequency domain, using a random-phase multisine signal to excite the system. The signal is designed such that the magnitude of its Fourier transform is constant in the frequency range \unit[0-10]{Hz}, allowing to excite all frequencies that are in the range of interest. As the phase shift of the sinusoids is drawn randomly, different time-domain realizations can be created, while preserving an identical amplitude spectrum in the frequency domain. This characteristic allows us to identify a standard deviation of the frequency response, $\hat{\sigma}_{\mathrm{nl}}(j\omega)$, which can be interpreted as a metric that captures the nonlinearity of the system.  

\begin{figure}
	\centering
	\includegraphics[trim={1.2cm 6.8cm 1.8cm 6.7cm},clip, scale=0.44]{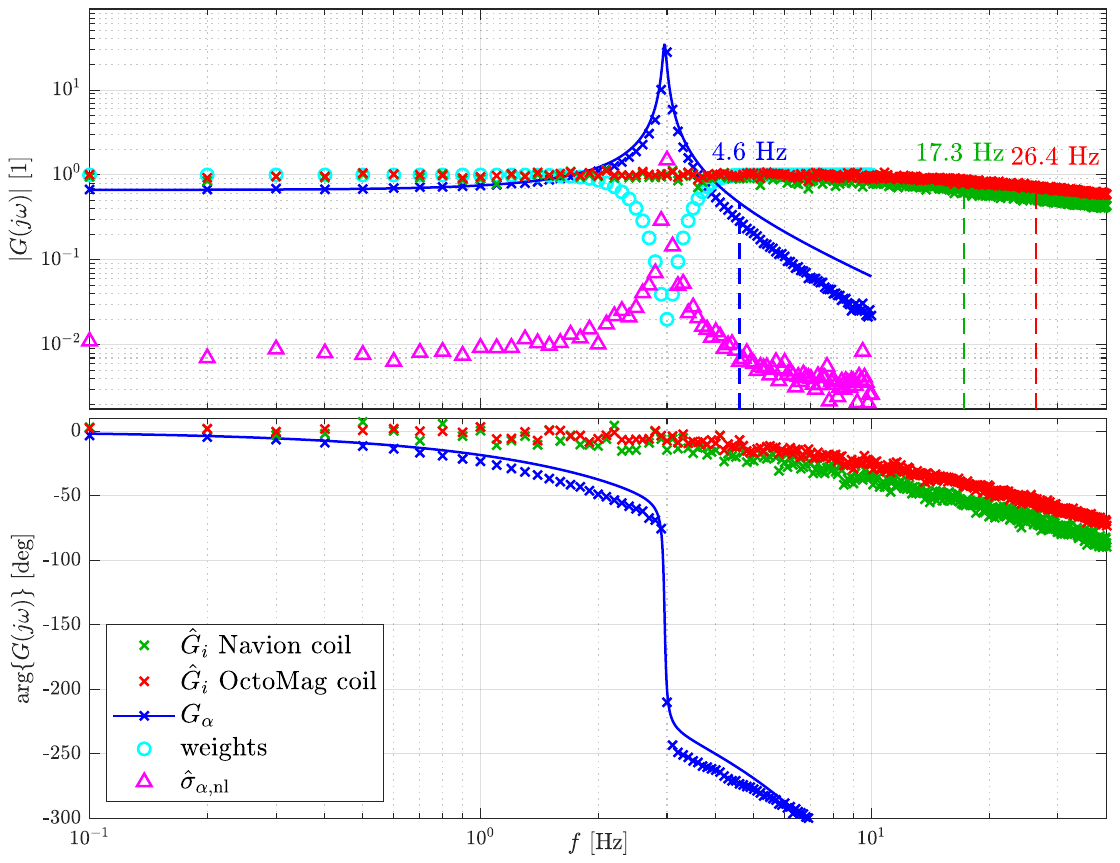} % left, bottom, right, top
	\vspace{-10pt}
	\caption{Bode diagram obtained from system identification experiments of the actuator dynamics (blue). For the actuator dynamics, similar results are observed for the $\beta$ direction. The standard deviation $\hat{\sigma}_{\alpha, \mathrm{nl}}$ captures the uncertainty of the estimate. We map $\hat{\sigma}_{\alpha, \mathrm{nl}}$ to values between $(0,1]$, used as weights (cyan) during curve fitting to obtain a parametric transfer function estimate (blue line). Furthermore, the dynamics of the closed-loop current control system for both the OctoMag coil (red) and the clinically ready Navion eMNS coil (green) are depicted; their respective bandwidths (\unit[-3]{dB} gain reduction) indicated by dashed lines. Although the bandwidth advantage is slightly reduced for the larger Navion coil, it still remains an order of magnitude higher than expected disturbances in potential clinical applications. }
	\label{fig:bode_multisine}
    \vspace{-4mm}
\end{figure}

We create $r = 10$ different excitation signals. To increase the signal-to-noise ratio, each excitation signal is repeated for ten consecutive periods, where the first four periods are discarded to minimize the influence of transients, resulting in $p = 6$ effective periods. Let $U_{\SP}^{[i, l]}(j \omega_k)$ and $Y^{[i, l]}(j \omega_k)$ denote the discrete Fourier transform of $u_\SP^{[i, l]}$ and $y^{[i, l]}$, which either represents $u_{\alpha, \SP}^{[i, l]}$ and $\alpha^{[i, l]}$ or $u_{\beta, \SP}^{[i, l]}$ and $\beta^{[i, l]}$, respectively. The index $i = 1, \ldots, p$ represents the index of the period within excitation signal $l$, with $l = 1, \ldots, r$. The averaged Fourier transform for excitation signal $l$ is then given by
\begin{align}
	Y^{[l]}(j \omega_k) = \frac{1}{p} \sum_{i = 1}^{p} Y^{[i,l]}(j \omega_k).
\end{align}  
Notice that $U_{\SP}^{[l]}(j \omega_k) = U_{\SP}^{[i,l]}(j \omega_k),\, \forall i = 1,\ldots,p$, as the Fourier transform is applied to the (noise-free) input set-point, $u_{\SP}$, which is identical for all periods within signal $l$. The empirical transfer function estimate for excitation signal $l$ and its average are then given by 
\begin{align*}
	\hat{G}^{[l]}(j\omega_k) = \frac{Y^{[l]}(j \omega_k)}{U^{[l]}(j \omega_k)}, \qquad \hat{G}_{ \mathrm{BLA}}(j\omega_k) = \frac{1}{r} \sum_{l=1}^{r} \hat{G}^{[l]}(j\omega_k), 
\end{align*}
where $\hat{G}_{\mathrm{BLA}}$ denotes the best linear approximator \cite{2012sysID_freqDomainAppr}. The sample variance $\hat{\sigma}^2_{\mathrm{nl}}$ reads as
\begin{align*}
	\hat{\sigma}_{\mathrm{nl}}^2\left(j \omega_k\right)=\frac{1}{r(r-1)} \sum_{l=1}^r\left|\hat{G}^{[l]}\left(j \omega_k\right)-\hat{G}_{\mathrm{BLA}}\left(j \omega_k\right)\right|^2.
\end{align*}
In Fig. \ref{fig:bode_multisine} the identified frequency response of the system, $\hat{G}_{\mathrm{BLA}}$, is depicted along with its uncertainty, $\hat{\sigma}_{\mathrm{nl}}$. Note that around the resonance frequency, the uncertainty associated with the estimate is elevated. We use this information to derive a frequency-dependent weight, $W(\omega_k)$, for the curve-fitting process. Specifically, in regions of small uncertainty, the values of $\hat{\sigma}_{\mathrm{nl}}$ are mapped to weights close to 1, while in high uncertainty zones, they approach a value close to 0, as shown in Fig. \ref{fig:bode_multisine}. We fit the parametric model 
\begin{align}
	G(s) = e^{-sT} \frac{b_0}{s^2 + a_1s + a_0} \label{eq:tf_fit}
\end{align} 
using the \texttt{tfest} function in Matlab\textregistered. Notice that we restrict the model fit to a second order model such that we can map the parameters, $a_1, b_0$, to the actuator dynamics of \eqref{eq:lin_eom} (with $M=0$, i.e. pendulum unattached), which are directly related to the parameters $d$ and $\mb$.

\subsection{Electrical Dynamics}
Employing the system identification methodology outlined in the preceding section, we analyze the dynamics $G_i$ of the closed-loop current control system. In particular, we examine the transfer function from the current setpoint, $i_{\text{SP}}$, to the measured current, $i$. This analysis is conducted on one of the OctoMag coils at an amplitude of \unit[5]{A} and a frequency range of \unit[0-40]{Hz}. Such an investigation allows us to examine the dynamic capabilities of eMNSs, facilitating an understanding of the frequency range within which effective disturbance rejection and trajectory tracking are feasible. The frequency response is depicted in Fig. \ref{fig:bode_multisine}. 

Additionally, using the same electrical driver, we perform the system identification experiment on a clinically scaled eMNS system known as Navion \cite{simone2024navion}. This comparison serves to evaluate whether the dynamic control strategies developed in this work can be extended to clinical-scale devices, which require larger coil volumes. Transitioning from the OctoMag coil to the Navion coil reduces the bandwidth from 26.4 to \unit[17.3]{Hz}. This frequency can be interpreted as the maximum frequency at which reliable field tracking and torque generation can occur. Control theory guidelines suggest that the actuation bandwidth should exceed the highest system frequency by a minimum factor of three to five, and for systems with unstable modes, by an order of magnitude. Considering disturbances at around \unit[3]{Hz} (equivalent to \unit[180]{bpm}), such as those that may occur during cardiac ablation in the human heart, these findings indicate that eMNSs have the potential for robust control in clinical settings. However, it is important to emphasize that each device (e.g. a catheter) exhibits unique dynamic properties and must be analyzed on a case-by-case basis. Nonetheless, the bandwidth of the eMNS remains a critical parameter, as it provides an upper limit on the achievable control performance in electromagnetic navigation. Notice, however, that the bandwidth is a function of the applied current amplitude and the specific driver hardware used. 

For the inverted pendulum considered in this work, the unstable eigenvalue derived from \eqref{eq:lin_eom} is approximately \unit[1.0]{Hz} for a \unit[40]{cm} pendulum and \unit[1.4]{Hz} for a \unit[20]{cm} pendulum. Generally, shorter pendula are more challenging to stabilize because their unstable mode has a smaller time constant. Note, however, that delay also plays an important role in determining the achievable control performance. Using \eqref{eq:tf_fit}, the delay is estimated to be about \unit[50]{ms}, which makes the \unit[20]{cm} pendulum particularly sensitive to disturbances (see video attachment). For reference, manually stabilizing a \unit[20]{cm} or even a \unit[30]{cm} pendulum exceeds the capabilities of most humans.

\section{State-Feedback Control}
\label{sec:State_Feedback_Control}
In this section, we discuss the implementation of a feedback controller stabilizing the pendulum in upright position. A prefilter is designed for reducing steady-state errors when tracking non-zero setpoints. Furthermore, we provide online compensation methods to errors resulting from the calibration of the magnetic field and the motion capture system. 

\subsection{Controller Design}
We discretize the continuous-time dynamics in \eqref{eq:state_space_cont_time} with a sampling time of $T_\mathrm{s} = \unit[10]{ms}$ using zero-order hold discretization. We denote the discrete-time state space model as 
% \begin{align}
% 	\x[k+1] &= \A \x[k] + \B u[k] \label{eq:state_space_discr_time} \\
% 	\bm{y}[k] &= \C \x[k],\,\, \text{with}\,\, \C = \< \I^{2x2} & \Zero^{2x2} \>,
% \end{align}
\begin{align}
	\x[k+1] &= \A \x[k] + \B u[k], \qquad\bm{y}[k] = \C\x[k], \label{eq:state_space_discr_time}
\end{align}
where, for the sake of brevity, $\A, \B, \C$ represents either $\A_\alpha, \B_\alpha, \C_\alpha$ or $\A_\beta, \B_\beta, \C_\beta$, with $\C = \< \I^{2 \times 2} & \Zero \>$. For each plane $(\alpha, \varphi)$, $(\beta, \theta)$ (the linearized dynamics are decoupled) we design a Linear Quadratic Regulator (LQR) that stabilizes the system around its upright equilibrium. Furthermore, the system can be stabilized around non-zero set-points, $\x_{\alpha, \SP}, \x_{\beta, \SP}$, provided that they satisfy the (dynamic) equilibrium conditions and remain near the system's upright equilibrium. The feedback policy for each plane, at time step $k$, is given by
% \begin{align}
% 	\< u_\alpha[k] \\ u_\beta[k] \> = \< \K_\alpha & \Zero \\ \Zero &  \K_\beta \> \< \x_{\alpha, \SP}[k] - \x_\alpha[k] \\ \x_{\beta, \SP}[k] - \x_\beta[k] \>,\label{eq:control_policy}
% \end{align}
\begin{align}
	u[k] = \K  (\x_{\SP}[k] - \x[k]), \label{eq:control_policy}
\end{align}
where we calculate the angular velocities in $\x_\alpha$ and $\x_\beta$ using finite differences. The gain matrices, $\K_\alpha, \K_\beta$, are obtained from the associated discrete-time algebraic Riccati equation\footnote{For the LQR design, we normalize the states and inputs as $\bar{\x} = \bm{T}_{\mathrm{x}}^{-1} \x$ and $\bar{u} = T_{\mathrm{u}}^{-1} u$. In particular, $\bm{T}_{\mathrm{x}} = \mathrm{diag}(\xi, \xi, 10\xi/s , 10\xi/s  ) \pi/180^\deg$ and $T_{\mathrm{u}} = \xi\pi/180^\deg$, where $\xi = 20^\deg$ is the maximum expected angle. We set $R = 10^3$ and choose $\bm{Q}=\mathrm{diag}(1.5, 1.5, 0.1, 0.1)$ for parameter case~i) and $\bm{Q}=\mathrm{diag}(0.8, 0.8, 0.1, 0.1)$ for parameter case~ii). Refining the LQR gains typically requires fewer than five iterations.}.

% plot successful stabilization of pendulum % 
% show capability of rejecting disturbance in phase (D)
% in same plot show effect of calibration error compensation 

% cause of the offset in analyzed in the two subsequent paragraphs

\subsection{Online Compensation of Steady-State Errors}
\begin{figure*}
	\begin{center}
		\includegraphics[trim={4.4cm 13.3cm 3cm 2.4cm},clip, scale=0.77]{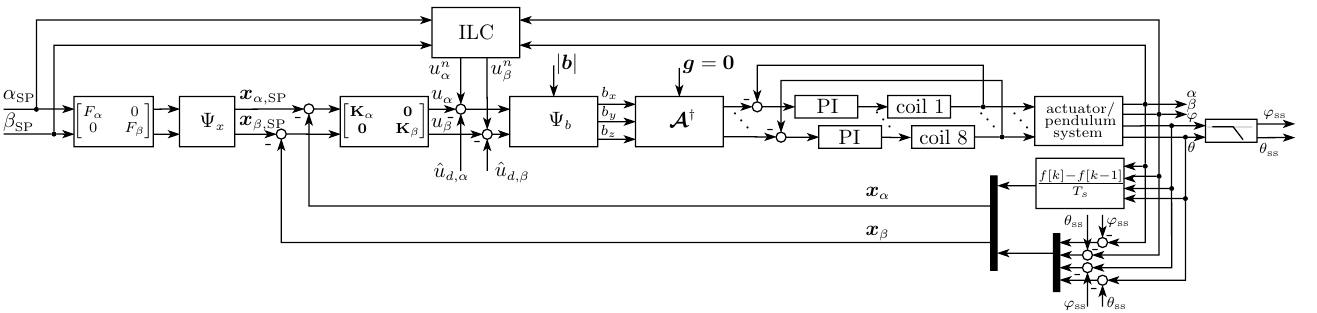} % left, bottom, right, top
	\end{center}
	\vspace{-0.5cm}
	\caption{Block diagram of the system illustrating the cascaded control structure. The prefilters, $F_\alpha, F_\beta$, are designed to reduce steady-state errors by scaling the setpoints appropriately. Setpoints, $\alpha_\SP, \beta_\SP$, are mapped to the state vector using $\mathrm{\Psi}_{x, \alpha}: \alpha_\SP \mapsto \< \alpha_\SP & 0 & \dot{\alpha}_\SP & 0 \>^\T$, and similarly for $\beta_\SP$. The state feedback controllers, $\K_\alpha, \K_\beta$, operate at \unit[100]{Hz} determining the magnetic field orientation, $u_\alpha, u_\beta$, which is converted to the magnetic field vector using the allocation $\PsiB$. The magnetic field vector is then converted into electrical currents using the actuation matrix, $\Act^\dag$. The electrical currents are controlled by eight independent PI controllers (drivers). Full state information is derived using finite-difference differentiation. An Iterative Learning Control (ILC) scheme is included calculating feedforward correction signals, $u_\alpha^n, u_\beta^n$, to counteract any repetitive error during the tracking of periodic reference trajectories.  }
	\label{fig:blockdiagram}
    \vspace{-5mm}
\end{figure*}

To ensure minimal steady-state errors while tracking non-zero set-points, a prefilter, $F$, is implemented. The main purpose of the prefilter is to scale the setpoints, as illustrated in the blockdiagram in Fig. \ref{fig:blockdiagram}, such that the errors are minimized when the system is in steady-state. 

It can be shown that only the actuator states, $\alpha, \beta$, are output controllable. Hence, we introduce the matrix, $\tilde{\C}_\alpha = \tilde{\C}_\beta = \< 1 & 0 & 0 & 0\>$, relating the state, $\x$, to the output controllable states, $\alpha$ and $\beta$. By substituting $\bm{y} = \bm{y_\SP}$ and applying the steady-state conditions $\x[k+1] = \x[k] = \x_\ss$ in \eqref{eq:state_space_discr_time}-\eqref{eq:control_policy}, the prefilter can be derived as
\begin{align}
	F = \biggl(\tilde{\C} \bar{\A}^{-1} \B \K \tilde{\C}^\T \biggr)^{-1}, \text{with}\,\,
	\bar{\A} \coloneqq \I - \A + \B\K,
\end{align}
where $F$ represents either $F_\alpha$ or $F_\beta$. 
 
An additional source of error results from minor misalignments that occur during the calibration of the motion capture system. The resulting misalignment between the $z$ axis of the inertial frame with the gravitational vector causes a significant steady-state error in the actuator angles. In \cite{hofer2023owc}, this misalignment is estimated in an online manner by low-pass filtering the pendulum angle, $\varphi_\ss$. By subtracting the learned steady-state offset, $\varphi_{\ss}$, from the measured angles, 
\begin{align}
	\alpha[k] \leftarrow \alpha[k] - \varphi_{\ss}[k], \qquad \varphi[k] \leftarrow \varphi[k] - \varphi_{\ss}[k]
	\label{eq:vicon_error_correction}
\end{align}      
we greatly reduce the actuator deflection error. 

In the process of determining the actuation matrix, $\Act$, minor inaccuracies can be encountered, which result in a small discrepancy between the desired and the actual magnetic field. We model this discrepancy as an additive disturbance, denoted as $u_d$, which impacts the control inputs, $u_\alpha$ and $u_\beta$. The steady-state relation from $u_d$ to $\x_\ss$ can be expressed as
\begin{align}
	\x_{\ss} = - \bar{\A}^{-1}\B u_{d} \,\,\Rightarrow\,\, \hat{u}_{d}[k] = -\bigl(\bar{\A}^{-1}\B\bigr)^\dag \x_{\ss}[k].
\end{align}     
This equation provides an estimation of the angular misalignment between the desired and the actual magnetic field vector, based on prior information of the identified model and real-time measurements. We can compensate for this misalignment by subtracting $\hat{u}_d[k]$ from the control input (see Fig. \ref{fig:blockdiagram}). Once converged, the learned offset is stored and remains fixed for the remainder of each experiment. The offset compensation methods described in this paragraph yield effects comparable to those of an integral controller, allowing to significantly reduce the steady-state error (see video attachment). However, in contrast to an integral controller the methods introduced here offer the benefit of delivering a distinct physical understanding.

%The controller is implemented using ROS \cite{quigley2009ROS} in Python on a conventional desktop computer (Intel Core i7 12-core CPU, 3.7 GHz), with the Linux Ubuntu 20.04 operating system.

\section{Iterative Learning Control}
\label{sec:Iterative_Learning_Control}
% mention potential application in medical (repetitive disturbances) ? -> source?
In this section, we present an ILC formulation that enables compensation of any repetitive errors that arise when tracking periodic reference trajectories while simultaneously balancing the pendulum. While the previously introduced state-feedback controller ensures the stabilization of the system and the rejection of (non-repeatable) disturbances, the ILC scheme presented in this section enhances the performance of the system by counteracting all disturbances that occur repetitively.  

ILC is recognized as a powerful tool in scenarios where tasks are repetitive, such as tracking periodic reference trajectories \cite{bristow2006ILC}. The underlying principle is the iterative improvement of the system's performance by leveraging error feedback from previous repetitions. We adopt a norm-optimal ILC formulation, wherein the learning computes a correction signal after the completion of each iteration by solving a quadratic optimization problem. This computation explicitly incorporates prior knowledge derived from our identified dynamic model. The design we propose is implemented in a parallel architecture, with the correction signal being added as a feedforward term to the state-feedback controller's output, as visualized in Fig. \ref{fig:blockdiagram}. Drawing upon the decoupled architecture of our state-feedback controller, it consequently follows that our ILC architecture is also decoupled. This results in two independent ILC schemes, running independently for the respective planes, $(\alpha, \varphi)$ and $(\beta, \theta)$. Despite its decoupled configuration, the ILC architecture maintains the ability to mitigate coupling effects, as the coupling dynamics become apparent through repetitive errors. Assuming one iteration has $N$ timesteps, we stack the control inputs, measurements, and setpoints of iteration $n$ in the following vectors, 
\begin{align*}
	\bm{u}_\alpha^n & :=\left[u_\alpha^n[0], \ldots, u_\alpha^n[N-1]\right]^\T \\
	\bm{y}_\alpha^n & :=\left[\alpha^n[0], \varphi^n[0], \ldots, \alpha^n[N-1], \varphi^n[N-1]\right]^\T \\
	\bm{y}_{\alpha, \SP} & :=\left[\alpha_{\SP}[0], \varphi_{\SP}[0], \ldots, \alpha_{\SP}[N-1], \varphi_{\SP}[N-1]\right]^\T, 
\end{align*}
and similarly for the $(\beta, \theta)$-plane. Here, $\bm{u}^n_\alpha$ is the correction signal and $\bm{y}_\alpha^n$ is the measured output within iteration $n$. The  setpoints, $\bm{y}_{\alpha, \SP}$, are identical for each iteration and hence are independent of the iteration index $n$. Note that $\varphi_{\SP}[k] = 0 \,\forall\, k = 0,\ldots, N-1$, which implies that we aim at keeping the pendulum in upright equilibrium, when tracking non-zero setpoints, $\alpha_\SP[k]$, with the actuator. For the sake of brevity, let $\bm{u}^n, \bm{y}^n, \bm{y}_\SP^n$ represent either $\bm{u}_\alpha^n, \bm{y}_\alpha^n, \bm{y}_{\alpha, \SP}^n$ or $\bm{u}_\beta^n, \bm{y}_\beta^n, \bm{y}_{\beta, \SP}^n$ in the following. As shown in \cite{bristow2006ILC}, the dynamics in the lifted state space can then be written as 
% \begin{align}
% 	\bm{y}^n &= \Pilc \bm{u}^n + \Pilc \bm{y}_\SP, \\
% 	\Pilc_{i, j} &= 
% 	\begin{cases} 
% 		\C (\A - \B\K)^{i-j} \B, & \text{if } i \geq j \\
% 		\Zero, & \text{otherwise},
% 	\end{cases}
% \end{align}
\begin{align}
	\bm{y}^n &= \Pilc \bm{u}^n + \Pilc \bm{y}_\SP, \\
	\Pilc_{i, j} &= \C (\A - \B\K)^{i-j} \B, \, \text{if } i \geq j ;\,\,\,\,\,
		\Pilc_{i, j}=\Zero, \, \text{otherwise}, \nonumber
\end{align}
where $i,j = 1,\ldots,N$. Let $\bm{e}^n \coloneqq \bm{y}_\SP - \bm{y}^n$ be the error in the $n$-th iteration. Inspired by \cite{zughaibi2021pickPlace}, we formulate the following objective function   
\begin{multline}
	J^{n+1}(\bm{u}^{n+1}) = {w_e \bm{e}^{n+1}}^\T \bm{e}^{n+1} + \\ (\bm{u}^{n+1} - \bm{u}^{n})^\T (\bm{u}^{n+1} - \bm{u}^{n}) + w_{\dot{u}} {\bm{u}^{n+1}}^\T \Dilc^\T \Dilc \bm{u}^{n+1}.  \nonumber
\end{multline} 
In this formulation, $w_e \geq 0$ and $w_{\dot{u}} \geq 0$ are scalar, fixed parameters. These parameters are used to weight the relative importance of the three main terms in the cost function with respect to each other. The first term penalizes the predicted error of the next iteration, the second term penalizes changes in the correction signal from iteration to iteration. The third term penalizes changes within the correction signal leveraging the derivative operator, $\Dilc \in \mathbb{R}^{N \times N}$, that approximates the derivative of $\bm{u}^{n+1}$ based on finite differences, defined as $\Dilc_{i,j} = -\delta_{i,j} + \delta_{i, j-1}, \forall i,j = 1,\ldots,N,$ with $\delta_{i,j}$ being the Kronecker delta. Since the optimization problem is unconstrained, we can derive the optimal solution in closed form:
\begin{align}
	{\bm{u}^{n + 1}}^* &= \Q \bm{u}^n + \L \bm{e}^n, \label{eq:ILC_update} \\
	\Q &= (w_e \Pilc^\T \Pilc + \I  + w_{\dot{u}} \Dilc^\T \Dilc)^{-1} ( w_e \Pilc^\T  \Pilc + \I ) \nonumber \\
	\L&= (w_e \Pilc^\T \Pilc + \I  + w_{\dot{u}} \Dilc^\T \Dilc)^{-1} \Pilc^\T w_e. \nonumber 
\end{align} 
Hence, at the end of each iteration, the correction signal for the next iteration, $\bm{u}^{n+1}$, is calculated by evaluating \eqref{eq:ILC_update}, with the precomputed matrices, $\Q \in \mathbb{R}^{N \times N}$, and $\L \in \mathbb{R}^{N \times 2N}$. 

\begin{figure}
	\centering
	\includegraphics[trim={17.1cm 8.20cm 17.5cm 10.0cm},clip, scale=0.31]{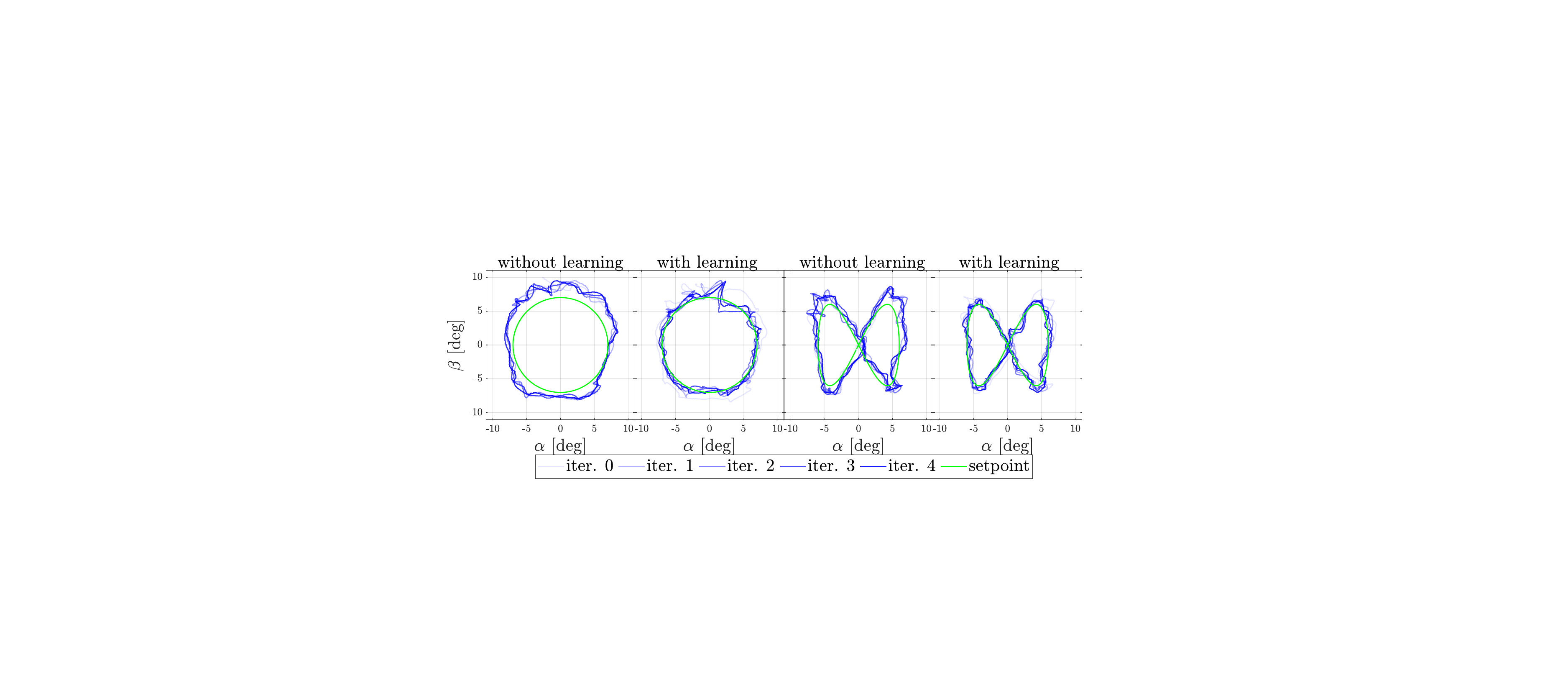} % left, bottom, right, top
	\includegraphics[trim={17.1cm 8.20cm 17.5cm 10.0cm},clip, scale=0.31]{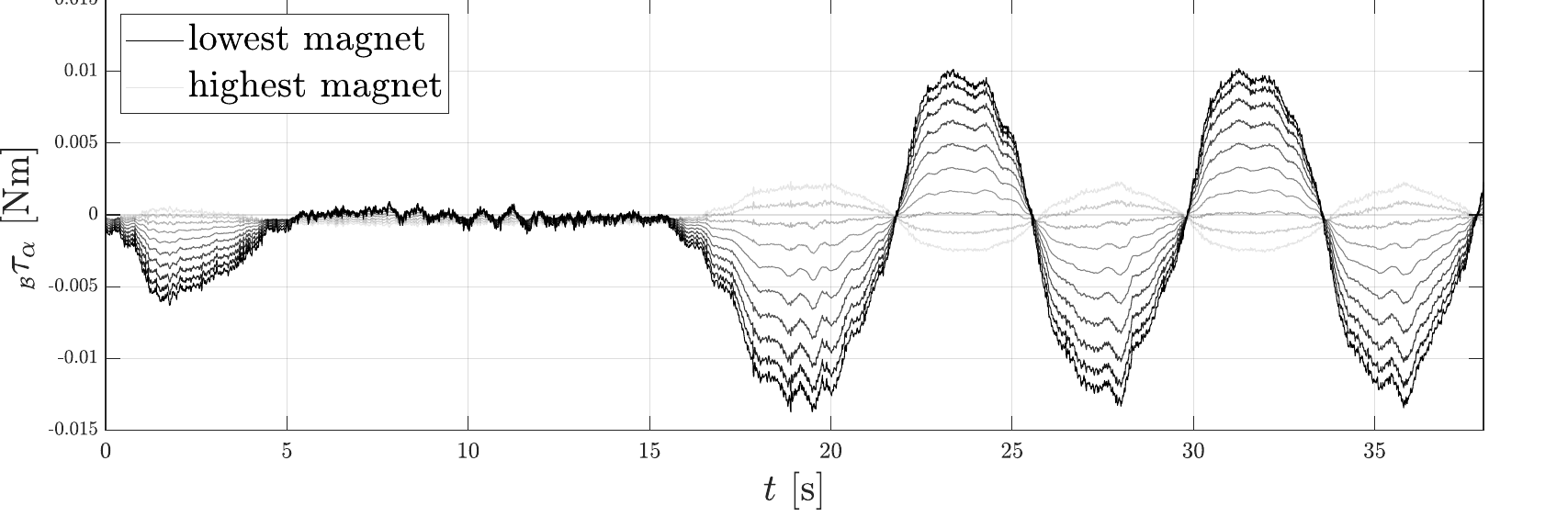} % left, bottom, right, top
    \vspace{-0.9cm}
	\caption{Experimental results tracking a circular and a figure eight trajectory, without learning and with activated learning control scheme (top plots). Iteration 0 corresponds to no learning. The ILC scheme is able to compensate for all repetitive disturbances. Due to the inherent instability of the inverted pendulum non-repetitve disturbances persist. The bottom plot shows the torque contribution (in body-fixed frame $\mathcal{B}$) for each of the ten magnets arising from the magnetic- and its gradient field, indicating that the magnetic field varies signifcantly over the magnetic volume, being the main source of the repetitve errors. Similar results are obtained for $\prescript{}{\mathcal{B}}{\tau_\beta}$.   }
	\label{fig:ILC_lissajous}
    \vspace{-5mm}
\end{figure}
In Fig. \ref{fig:ILC_lissajous}, we present the experimental results obtained after four iterations of learning with ILC. The proposed method effectively compensates for repetitive errors. Specifically, the root-mean-square error is reduced from 1.7\deg\,during the initial iteration (without learning) to 0.5\deg\,by the fourth iteration during the circular maneuver. The dominant source of these repetitive errors are unmodeled disturbances induced by magnetic field gradients. To assess their impact, we examine the net torque on the system, 
\begin{align}
   \bm{\tau} = |\tilde{M}| \int_V \biggl( \bm{n} \times \bm{b}(\bm{r}, \bm{i} ) + \bm{r} \times \frac{\partial \bm{b}(\bm{r}, \bm{i} )}{\partial \bm{r}} \bm{n} \biggr) \d V. \label{eq:net_torque}
\end{align}
In this expression, the first term in the integrand represents torques generated by the magnetic field, while the second term represents torques arising from the forces, which in turn are caused by the field gradients acting on the magnetic volume. Here, $|\tilde{M}|$ is the magnetization in \unit[]{A/m}, $\bm{n}$ is the dipole’s orientation, and both $\bm{b}(\bm{r}, \bm{i})$ and its gradient are determined using the nonlinear multipole expansion model described in \cite{petruska2017multipole}. We discretize \eqref{eq:net_torque} into ten elements, each corresponding to the torque contribution of a single magnet. As shown in Fig. \ref{fig:ILC_lissajous}, the top magnets produce a torque opposite to that of the bottom magnets, indicating significant spatial variation of the magnetic field. Notably, this suggests that stabilization would be more efficient with fewer magnets. Still, the ILC scheme can mitigate these gradient-induced errors to a certain extent. Its limitations arise from the inherent structure of the control input, which only influences the magnetic field vector. Exploiting the coil redundancy to explicitly control gradients could further improve performance.

\section{Conclusion}
\label{sec:Conclusion}

This letter highlights the dynamic capabilities of an eMNS by successfully stabilizing a 3D inverted pendulum. A dynamic model is identified, which is used to synthesize a feedback controller that stabilizes the inherently unstable pendulum system. An ILC scheme is implemented to improve trajectory tracking while simultaneously maintaining the pendulum in upright position. To explore the dynamic limits of eMNSs, we perform a comprehensive analysis of the electrical system dynamics and demonstrate that the high-bandwidth benefits are preserved in systems equipped with clinical-scale coils.    

The dynamic modeling and control strategies presented herein generalize to tasks well beyond balancing a pendulum, offering applicability for a range of magnetically controlled objects. The techniques successfully leverage the high actuation bandwidth of eMNSs. This is crucial for enabling new medical applications, such as eMNS-guided cardiac ablation, where precise and responsive control is essential. 

In this work, the direction of the magnetic field vector serves as the control input, whereas magnetic field gradients are treated as unmodelled disturbances. The proposed control architecture compensates for these disturbances during non-equilibrium trajectory tracking, albeit within a defined radius. Beyond this radius, the magnetic field gradients intensify excessively. Future research will explore employing magnetic field gradients as additional control inputs, thereby harnessing the full potential of all eight coils in the OctoMag system. 
%"a visual demonstration of dynamic capabilities of electromagnetic navigation systems"
\vspace{-1mm}
\section*{Acknowledgments}
The authors would like to thank Hao Ma for sharing his knowledge on system identification and the team of MagnebotiX AG for the technical support. Further thanks go to the Max Planck ETH Center for Learning Systems and the Swiss National Science Foundation for the financial support. Michael Muehlebach thanks the German Research Foundation and the Branco Weiss Fellowship, administered by ETH Zurich, for the financial support. 

\vspace{-1mm}
\section*{Conflict of Interests}
Bradley Nelson is co-founder of MagnebotiX AG, which commercializes the OctoMag system. The other authors declare no conflict of interest.  

% Appendix: 
%{\appendix[Actuation Matrix]

% provide actuation matrix? 

%}

\vspace{-1mm}
\bibliographystyle{IEEEtran}
\bibliography{IEEEabrv,bibliography_abbrv}

\vfill

\end{document}